\documentclass[preprint,showpacs,preprintnumbers,amsmath,amssymb]{revtex4}
 \usepackage{graphicx}% Include figure files
 \usepackage{dcolumn}% Align table columns on decimal point
 \usepackage{bm}% bold math

\begin{document}

\title{Neutron pair transfer in  sub-barrier capture process}
\author{
V.V.Sargsyan$^{1,2}$, G.~Scamps$^{3}$, G.G.Adamian$^{1}$, N.V.Antonenko$^1$, and  D.~Lacroix$^{4}$
}
\affiliation{
$^{1}$Joint Institute for Nuclear Research, 141980 Dubna, Russia\\
$^{2}$International Center for Advanced Studies, Yerevan State University, M. Manougian 1, 0025, Yerevan, Armenia\\
$^{3}$GANIL, 14076 Caen Cedex, France \\
$^{4}$Institut de Physique Nucl\'eaire, IN2P3-CNRS, Universit\'e Paris-Sud, F-91406 Orsay Cedex, France
}
\date{\today}

\begin{abstract}
The sub-barrier capture  reactions following the neutron pair transfer
are proposed to be used for the indirect study of neutron-neutron correlation
in the surface region of nucleus.
%The strong surface enhancement of the neutron pairing
The strong effect of the dineutron-like clusters transfer stemming from the surface of magic and non-magic nuclei $^{18}$O, $^{48}$Ca, $^{64}$Ni, $^{94,96}$Mo, $^{100,102,104}$Ru, $^{104,106,108}$Pd, and $^{112,114,116,118,120,124,132}$Sn is demonstrated.
The dominance of two-neutron transfer channel at the vicinity of the Coulomb barrier
%compared to other transfer channels
is further supported by time-dependent mean-field approaches.
%The energy range where the two-particle transfer dominates is shown
%to become wider as one of the nucleus becomes more exotic.
%It is found that at the vicinity of
% the Coulomb barrier the two-nucleon transfer dominates.
% This window becomes larger as the nuclei become more exotic.
%The dominance of the dineutron-like clusters
%in the surface of magic and non-magic nuclei
%$^{18}$O, $^{48}$Ca, $^{64}$Ni,
%$^{94,96}$Mo, $^{100,102,104}$Ru, $^{104,106,108}$Pd,
%and
%$^{112,114,116,118,120,124,132}$Sn is demonstrated.
\end{abstract}


 \maketitle

\section{Introduction}
%In the past a considerable number of experimental and theoretical studies of two-neutron transfer have been accumulated
%for the stable nuclei~\cite{vonOertzen}.
 Two-neutron transfer reactions such as ($p$,$t$) or ($t$,$p$) have been used for many years in order to study the
 nucleon pairing correlations in the stable nuclei~\cite{BohrNathan,vonOertzen}.
 The corresponding pair transfer modes are usually described in terms of
 pairing vibrations or pairing rotations~\cite{Broglia,Kart},
 which  are
associated with the pair correlation.
It has been established  that the two-neutron transfer amplitude
is influenced by collective modes caused by the Cooper-pair superfluidity~\cite{vonOertzen}.
%The relation between the characteristics of the pairing correlations and the transfer probabilities in
% two-particle transfer reactions has been analyzed.
% For example, the $2n$ ($p$,$t$) transfer reactions
% in very neutron-rich Sn  isotopes have been
% indicated as a good experimental candidate.
In the superfluid nuclei $^{18}$O, $^{206,210}$Pb, and $^{114}$Sn, the Cooper pair with short range space correlation
has been theoretically predicted~\cite{past}. The size of the Cooper pair
is estimated to be comparable to the average inter-nucleon distance~\cite{past}.

 Recently, there is a renewal of interest on
 experimental nucleon pair, alpha cluster,
%  $1n$, $2n$,
and more generally multinucleon
transfer channels at bombarding energies above and below the Coulomb
 barriers~\cite{Corradi,Corradi2,Simenel}.
%L. Corradi, A. M. Stefanini, D. Ackermann, S. Beghini, G. Montagnoli,
%C. Petrache, F. Scarlassara, C. H. Dasso, G. Pollarolo, and A. Winther
%Phys. Rev. C {\bf 49}, R2875 (1994);
%L. Corradi {\it at al.},
%Phys. Rev. C {\bf 54}, 201 (1996)
%L. Corradi, A. M. Stefanini, J. H. He, S. Beghini, G. Montagnoli, F. Scarlassara,
%G. F. Segato, G. Pollarolo, and C. H. Dasso
%Phys. Rev. C 56}, 938 (1997);
%L. Corradi, A. M. Stefanini, C. J. Lin, S. Beghini, G. Montagnoli, F. Scarlassara, G. Pollarolo, and A. Winther
%Phys. Rev. C {\bf 59}, 261 (1999);
%L. Corradi {\it at al.},
%Phys. Rev. C {\bf 61}, 024609 (2000);
%L. Corradi {\it at al.},
%Phys. Rev. C {\bf 63}, 021601 (2001);
%S. Szilner {\it at al.},
%Phys. Rev. C {\bf 71}, 044610 (2005);
%S. Szilner {\it at al.},
%Phys. Rev. C {\bf 76}, 024604 (2007);
%L.~Corradi {\it at al.},
%Phys. Rev. C {\bf 84}, 034603 (2011);
%M. Evers, M. Dasgupta, D.J. Hinde, D.H. Luong, R. Rafiei, R. du Rietz, and C. Simenel,
%Phys. Rev. C {\bf 84}, 054614 (2011). ].
The effect of  correlations between nucleons on the nuclear breakup or decay mechanism has been
studied both experimentally and theoretically~\cite{Marques,Kolata,Ershov,Lacroix0,Grigor,Spyrou}.
Studies of pairing effects in both finite nuclei and
nuclear matter have  intensified interests in the
recent  years~\cite{Fortunato,Volya,Dean,Khan,Saper1,Matsuo,Grasso,Broglia2,Lacroix,Lacroix2,Lacroix3,Sambataro}.
Attention has been paid to the properties of the pair correlation in the neutron-rich nuclei
with the neutron skin and the neutron halo~\cite{Zhukov,Mar,Barranco,Mat}.
The ($p$,$t$) reactions on light-mass
neutron-rich nuclei such as $^{6,8}$He  and $^{11}$Li
point out the importance of the pair correlations
in these typical halo or skin nuclei.
%A spatial two-neutron correlation or the di-neutron correlation, between
%two weakly bound neutrons forming the halo in $^{6,8}$He and $^{11}$Li, has been discussed extensively.
The experimental signatures
of a spatial two-neutron correlation or the di-neutron correlation  between
two weakly bound neutrons forming the halo in $^{6,8}$He and $^{11}$Li
have been reported  in Refs.~\cite{Ter-Akopian,DeYoung,Nakamura,Moeller,Chatterjee}.
There exists also several studies demonstrating enhancement of the pair correlation
in the nuclear surface and exterior regions of the
neutron-rich nuclei~\cite{Khan,Saper1,Matsuo,Grasso,Lacroix,Saper2,Hag05}.
%The relation between the characteristics of the pairing correlations and the transfer probabilities in
%two-particle transfer reactions has been analyzed~\cite{Khan,Matsuo,Grasso,Lacroix}.
%For example, the $2n$ ($p$,$t$) transfer reactions
% in very neutron-rich Sn  isotopes have been
%indicated as a good experimental candidate.
%
%The giant pairing vibration in neutron-rich nuclei has also been discussed~\cite{Fortunato,Avez}.
%In the case of the giant pairing vibrations the excitation involves the promotion of a pair
%of particles (or holes) in the next major shell (hence  in excitation energy around 2$\hbar\omega$)
%and is expected to display a collective pairing strength comparable with low-lying vibrations.
%
%The pair transfer feeding the excited 0$^+$ and first 2$^+$ states in the neutron-rich nuclei
%has been investigated  from the viewpoint of the pair vibrational mode.
%, which  is one of collective vibrational modes associated with the pair correlation.
A possible link between the pair transfer
%between 0$^+$ ground states
and the
surface enhancement of the pairing in medium and heavy neutron-rich nuclei has been suggested in
Ref.~\cite{Dobaczewski} and more recently discussed in \cite{Matsuo,Broglia2,Grasso,Lacroix,Gra13}
It has been
argued in Ref.~\cite{Khan} that the pair transfer can be a possible probe of different models of the pairing interaction.
In literature~\cite{Schuck}, the origin of the small size of Cooper pair on the nuclear surface
%(a consequence of the enhanced pairing correlations or of the finiteness of the single-particle wave functions)
is still under discussions. It can be a consequence of the enhanced pairing correlations or of the finiteness
of the single-particle wave functions.
%

%Thus, in a finite many-body system, like the nucleus, the configuration mixing caused by
% the pairing interaction leads to
A strong spatial correlation between the nucleons  gives rise to specific  features
like dineutron or alpha clustering formation and to the
possibility of a contribution to the transfer from the simultaneous
one-step pair   transfer mechanism.
By describing the capture (fusion) reactions at sub-barrier energies within the quantum diffusion approach,
we want to demonstrate indirectly the strong
%surface enhancement of the neutron pairing
dineutron spatial correlations
in the surface region of
stable  nuclei.
We will consider the capture reactions with  the negative one-neutron transfer  ($Q_{1n}<0$)
and the positive two-neutron transfer ($Q_{2n}>0$) (before  crossing  the Coulomb barrier), where
the one-step neutron pair transfer  is expected to be dominant.
The study of this process is one of the important points in the understanding of pairing
correlations in nuclei. The distinction between two-step sequential and  one-step cluster transfer
is a great challenge, not only in nuclear physics but also in electron transfer between ions or atomic cluster
collisions~\cite{vonOertzen}.
%In the extreme case of infinitely strong correlation the one-step mechanism is expected to prevail,
%the two-step sequential process being instead dominant in the case of pure uncorrelated nucleons.
Note that the capture (fusion) reaction following the neutron pair transfer  is
the  indirect way of the study of pairing effects.

\section{Model}
In the quantum diffusion approach~\cite{EPJSub,EPJSub1,EPJSub2,EPJSub3}
the collisions of  nuclei are treated in terms
of a single collective variable: the relative distance  between
the colliding nuclei. The  nuclear deformation effects are taken into consideration through the dependence
of the nucleus-nucleus potential on the deformations and mutual  orientations of the colliding nuclei.
Our approach takes into account the fluctuation and dissipation effects in
the collisions of heavy ions which model the coupling with various channels
(for example, coupling of the relative motion with the non-collective single-particle excitations and
low-lying collective modes
such as dynamical quadrupole and octupole excitations of the target and projectile~\cite{Ayik333}).
We have to mention that many quantum-mechanical and non-Markovian effects accompanying
the passage through the potential barrier are considered in our formalism~\cite{EPJSub,our}
through the friction and diffusion. The two-neutron
transfer with the  positive  $Q_{2n}$-value was taken into consideration in~\cite{EPJSub,EPJSub2}.
Our assumption is that, just before the projectile is captured by the target-nucleus
(i.e. just before the crossing of the Coulomb barrier),
the two-neutron  transfer  occurs   and can lead to the
population of the first excited collective state in the recipient
nucleus~\cite{Corradi2,SSzilner}
(the donor nucleus remains in the ground state).
So, the motion to the
$N/Z$ equilibrium starts in the system before the capture
because it is energetically favorable in the dinuclear system in the vicinity of the Coulomb barrier.
For the reactions under consideration,
the average change of mass asymmetry is connected to the  two-neutron
transfer ($2n$-transfer).
Since after the transfer the mass numbers, the isotopic composition and the deformation parameters
of the interacting nuclei, and, correspondingly, the height $V_b=V(R_b)$
[$R=R_b$ is the position of the Coulomb barrier]
%($R_b$ is the position of the Coulomb barrier) are changed,
and shape of the Coulomb barrier change,
one can expect an enhancement or suppression  of the capture.
If  after the neutron transfer the deformations of interacting nuclei increase (decrease),
the capture probability increases (decreases).
When the isotopic dependence of the nucleus-nucleus
potential is weak and   after the transfer the deformations of interacting nuclei do not change,
there is no effect of the neutron transfer on the capture.
In comparison with Ref. \cite{Dasso}, we assume that the negative transfer $Q-$values
do not play  visible  role in the capture process.
Our scenario was verified in the description of many reactions \cite{EPJSub2}.
The calculated results for all reactions
are obtained with the same set of parameters as in Refs. \cite{EPJSub1,EPJSub2}
and are rather insensitive
to the reasonable variation of them.
One should note that the diffusion models, which  include  quantum statistical effects,
were also treated in Refs. \cite{Hofman,Ayik,Hupin}.

%The heights of the calculated Coulomb barriers $V_b=V(R_b)$
%($R_b$ is the position of the Coulomb barrier)
%are adjusted to the experimental data
%for the fusion or capture cross sections.

The capture cross section is the sum of the partial capture cross sections \cite{EPJSub,EPJSub1,EPJSub2}
\begin{widetext}
\begin{eqnarray}
\sigma_{cap}(E_{\rm c.m.})&=&\sum_{J}^{}\sigma_{\rm cap}(E_{\rm
c.m.},J)=\nonumber\\&=& \pi\lambdabar^2
\sum_{J}^{}(2J+1)\int_0^{\pi/2}d\theta_1\sin(\theta_1)\int_0^{\pi/2}d\theta_2\sin(\theta_2) P_{\rm cap}(E_{\rm
c.m.},J,\theta_1,\theta_2),
\label{1a_eq}
\end{eqnarray}
\end{widetext}
where $\lambdabar^2=\hbar^2/(2\mu E_{\rm c.m.})$ is the reduced de Broglie wavelength,
$\mu=m_0A_1A_2/(A_1+A_2)$ is the reduced mass ($m_0$ is the nucleon mass),
and the summation is over the possible values of the angular momentum $J$
at a given bombarding energy $E_{\rm c.m.}$.
Knowing the potential of the interacting nuclei for each orientation with the angles $\theta_i (i=1,2)$, one can obtain the partial capture probability
$P_{\rm cap}$ which is defined by the probability to penetrate the potential barrier in the relative distance coordinate $R$
 at a given $J$.
The value of $P_{\rm cap}$
is obtained by integrating the propagator $G$ from the initial
state $(R_0,P_0)$ at time $t=0$ to the final state $(R,P)$ at time $t$ ($P$ is the momentum):
\begin{eqnarray}
P_{\rm cap}&=&\lim_{t\to\infty}\int_{-\infty}^{r_{\rm in}}dR\int_{-\infty}^{\infty}dP\  G(R,P,t|R_0,P_0,0)\nonumber \\
&=&\lim_{t\to\infty}\frac{1}{2} {\rm erfc}\left[\frac{-r_{\rm in}+\overline{R(t)}}
{{\sqrt{\Sigma_{RR}(t)}}}\right].
\label{1ab_eq}
\end{eqnarray}
Here, $r_{\rm in}$ is an internal turning point.
The second line in (\ref{1ab_eq}) is obtained by using the propagator
$G=\pi^{-1}|\det {\bf \Sigma}^{-1}|^{1/2}
\exp(-{\bf q}^{T}{\bf \Sigma}^{-1}{\bm q})$
(${\bf q}^{T}=[q_R,q_P]$,
$q_R(t)=R-\overline{R(t)}$, $q_P(t)=P-\overline{P(t)}$, $\overline{R(t=0)}=R_0$,
$\overline{P(t=0)}=P_0$, $\Sigma_{kk'}(t)=2\overline{q_k(t)q_{k'}(t)}$, $\Sigma_{kk'}(t=0)=0$,
$k,k'=R,P$) calculated  for
an inverted oscillator which approximates
the nucleus-nucleus potential $V$ in the variable $R$.
At given $E_{\rm c.m.}$ and $J$, the
classical action is calculated for the realistic nucleus-nucleus potential.
Then the   realistic nucleus-nucleus potential is replaced by an inverted oscillator
which has the same barrier height and classical action. So, the frequency
 $\omega(E_{\rm c.m.},J)$ of
this oscillator is set to obtain an equality of the
classical actions in the approximated and realistic potentials.
The action is calculated in the WKB approximation which is the accurate
at the sub-barrier energies.
Usually in the literature the parabolic approximation with
$E_{\rm c.m.}$-independent $\omega$ is employed
which is not accurate at the deep sub-barrier energies.
%Note that the passage through the Coulomb barrier approximated by a parabola
%has been previously studied in Ref. \cite{Hofman}.
Our approximation is well justified for the
reactions and energy range considered here \cite{EPJSub,EPJSub1,EPJSub2}.
Finally, one can find the expression for the  capture probability:
\begin{eqnarray}
P_{\rm cap}&=&
\frac{1}{2} {\rm erfc}\left[\left(\frac{\pi s_1(\gamma-s_1)}{2\hbar\mu(\omega_0^2-s_1^2)}\right)^{1/2}
\frac{\mu\omega_0^2 R_0/s_1+P_0}
{\left[\gamma \ln(\gamma/s_1)\right]^{1/2}}\right],
\label{PC_eq}
\end{eqnarray}
where $\gamma$ is the internal-excitation width,  $\omega_0^2=\omega^2\{1-\hbar\tilde\lambda\gamma/
[\mu(s_1+\gamma)(s_2+\gamma)]\}$ is  the renormalized frequency in the Markovian limit,
the value of $\tilde\lambda$ is related to the strength of linear coupling
in the coordinates between collective and internal subsystems.
The non-Markovian effects appear in the calculations
through  $\gamma$. Here, $\hbar\gamma$=15 MeV.
 The
$s_i$ are the real roots ($s_1\ge 0> s_2 \ge s_3$) of the following equation~\cite{EPJSub,EPJSub1,EPJSub2}:
\begin{eqnarray}
(s+\gamma)(s^2- \omega_0^2)+\hbar\tilde\lambda\gamma s/\mu=0.
\label{Root_eq}
\end{eqnarray}
%The details of the used formalism are presented in~\cite{EPJSub,EPJSub1}.
%We have to mention that most of the quantum-mechanical, dissipative  effects and non-Markovian effects accompanying
%the passage through the potential barrier are taken into consideration in our formalism \cite{EPJSub,EPJSub1,our}.
%For example, the non-Markovian effects appear in the calculations
%through the internal-excitation width $\gamma$.

As  shown in Refs.~\cite{EPJSub,EPJSub1}, the nuclear forces start to play a role
at $R_{int}=R_b+1.1$ fm where the nucleon density of the colliding nuclei approximately reaches
10\% of the saturation density.
If the value of $r_{\rm ex}$ corresponding to the external turning point
is larger than the interaction radius $R_{int}$,
we take $R_0=r_{\rm ex}$ and $P_0=0$ in Eq.~(\ref{PC_eq}).
For $r_{\rm ex}< R_{int}$, it is natural
to start our treatment with $R_0=R_{int}$ and $P_0$ defined by the kinetic energy
at $R=R_0$. In this case the friction hinders the classical motion to proceed towards smaller values of $R$.
If $P_0=0$ at $R_0>R_{int}$, the friction almost does not play a role in the transition
through the barrier.
Thus, two regimes of interaction at sub-barrier energies differ by the action
of the nuclear forces and the role of friction at $R=r_{\rm ex}$.

To calculate the nucleus-nucleus interaction potential $V(R)$,
we use the procedure described in Refs.~\cite{EPJSub,EPJSub1,EPJSub2,AAShobzor}.
For the nuclear part of the nucleus-nucleus potential, the double-folding formalism with
the Skyrme-type density-dependent effective nucleon-nucleon interaction is used.
The parameters of the potential were adjusted to describe the experimental
data at energies above the Coulomb barrier corresponding to spherical nuclei.
The absolute values of the quadrupole deformation parameters $\beta_2$ of even-even deformed nuclei
and of the first excited collective states of nuclei were taken from Ref.~\cite{Ram}.
For the  nuclei deformed in the
ground state, the $\beta_2$ in the first excited collective state is similar
to the $\beta_2$ in the ground state.
For the double magic  nuclei,
we take $\beta_2=0$ in the ground state.
For the rest of nuclei, we used
 the ground-state
quadrupole deformation parameters  extracted in Ref. \cite{EPJSub2}
from a comparison
of the calculated capture cross sections with the existing experimental data.

%\section{Results of calculations}
\section{Influence of neutron pair transfer
on  capture}
The choice of the projectile-target combination is crucial in the understanding of pair transfer phenomenon
in the capture process. In the capture reactions with
$Q_{1n}<0$ and $Q_{2n}>0$, the two-step sequential transfer  is almost closed before capture.
So, choosing properly  the reaction combination, one can reduce the  successive
transfer  in the process. For the systems studied, one can make unambiguous statements
regarding the neutron transfer process with a positive $Q_{2n}$ value when the interacting
nuclei are double magic or semimagic nuclei. In this case one can disregard the strong
nuclear deformation effects before the neutron transfer.
\begin{figure}[tbp]
\includegraphics[scale=1]{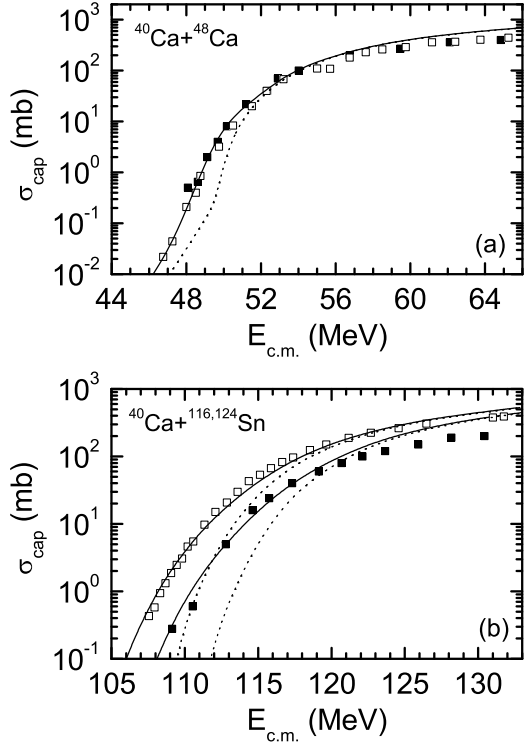}
\caption{The calculated (lines) and experimental
(symbols)~\protect\cite{trotta40ca48ca,Stefanini40ca116124sn}
capture cross sections
vs
$E_{\rm c.m.}$ for the  reactions $^{40}$Ca+$^{48}$Ca (a) and
$^{40}$Ca+$^{116,124}$Sn (b).
The calculated capture cross sections without taking into account
the neutron pair transfer  are shown by dotted lines.
}
\label{1_fig}
\end{figure}
\begin{figure}[tbp]
\includegraphics[scale=1]{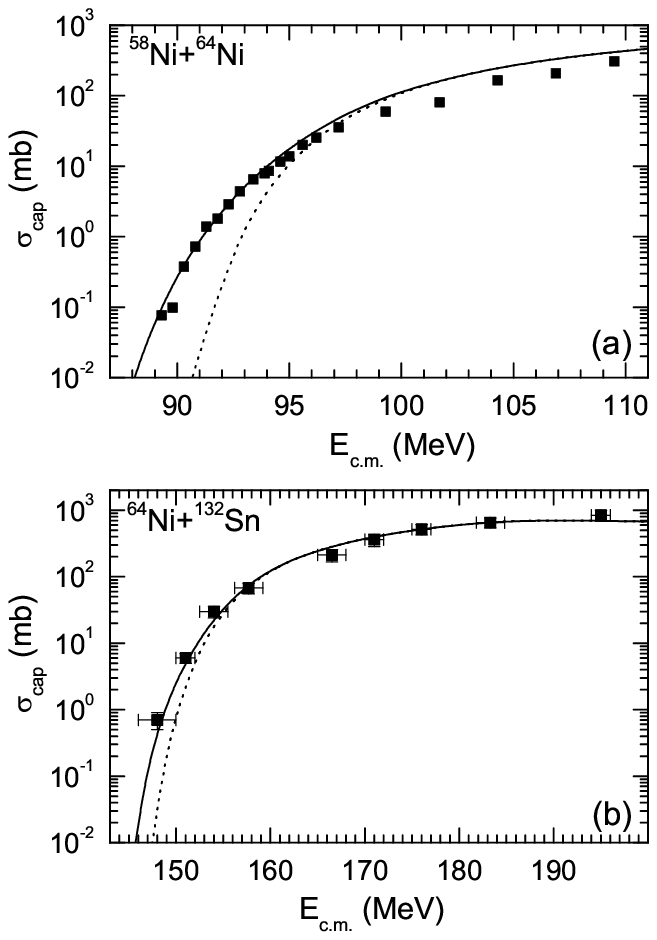}
\caption{The same as in Fig.~1, but for the  reactions $^{58}$Ni+$^{64}$Ni (a) and
$^{64}$Ni+$^{132}$Sn (b). The experimental data are from
Refs.~\protect\cite{Beckerman58Ni5864Ni74Ge,LiangNi64Sn132}.
}
\label{2_fig}
\end{figure}

In Figs.~1 and 2 the calculated capture cross sections for the reactions
$^{40}$Ca + $^{48}$Ca  ($Q_{1n}=-1.6$  MeV,  $Q_{2n}=2.6$  MeV),
$^{40}$Ca + $^{116}$Sn ($Q_{1n}=-1.2$  MeV,  $Q_{2n}=2.8$  MeV),
$^{40}$Ca + $^{124}$Sn ($Q_{1n}=-0.1$  MeV,  $Q_{2n}=5.4$  MeV),
$^{58}$Ni + $^{64}$Ni  ($Q_{1n}=-0.66$ MeV,  $Q_{2n}=3.9$  MeV),
and
$^{64}$Ni + $^{132}$Sn ($Q_{1n}=-1.21$ MeV,  $Q_{2n}=2.5$  MeV)
are in a good agreement with the available experimental
data~\cite{trotta40ca48ca,Stefanini40ca116124sn,Beckerman58Ni5864Ni74Ge,LiangNi64Sn132}.
In all reactions $1n$-neutron transfer  is closed ($Q_{1n}<0$) and
$Q_{2n}$-values for the $2n$-transfer
processes are positive.
%there are $2n$-neutron transfer process with positive $Q_{2n}$-values.
%The $Q_{2n}$-values for the $2n$-transfer
%processes are positive  for all reactions.
Thus, the $2n$-neutron transfer is more important  for a
good description of the experimental data than the $1n$-neutron transfer.
The influence  of the $2n$-neutron transfer on the
capture cross section occurs due to the change of the isotopic composition
and the deformations of  the reaction partners.
The   $2n$-transfer indirectly influence the quadrupole
deformation of the nuclei.
When after the neutron transfer (just before the crossing of the Coulomb barrier)
in the reactions
$^{40}$Ca($\beta_2=0$)+$^{48}$Ca($\beta_2=0$)$\to ^{42}$Ca($\beta_2=0.247$)+$^{46}$Ca($\beta_2=0$),
$^{40}$Ca($\beta_2=0$)+$^{116}$Sn($\beta_2=0.112$)$\to ^{42}$Ca($\beta_2=0.247$)+$^{114}$Sn($\beta_2=0.121$),
$^{40}$Ca($\beta_2=0$)+$^{124}$Sn($\beta_2=0.095$)$\to ^{42}$Ca($\beta_2=0.247$)+$^{122}$Sn($\beta_2=0.104$),
$^{58}$Ni($\beta_2=0.05$)+$^{64}$Ni($\beta_2=0.087$)$\to ^{60}$Ni($\beta_2=0.207$)+$^{62}$Ni($\beta_2=0.087$),
and
$^{64}$Ni($\beta_2=0.087$)+$^{132}$Sn($\beta_2=0$)$\to ^{66}$Ni($\beta_2=0.158$)+$^{130}$Sn($\beta_2=0$)
the deformations of nuclei
increase, the values of the corresponding Coulomb barriers  decrease.
As a result, the two-neutron  transfer
enhances the capture process in these reactions at the sub-barrier energies.
The enhancement becomes  stronger with decreasing bombarding energy  (Figs.~1 and 2).
Previously, the importance of the neutron pair
transfer in the capture (fusion) process was stressed in Refs.~\cite{Dasso,Pengo,Stef}.
\begin{figure}[tbp]
\includegraphics[scale=1]{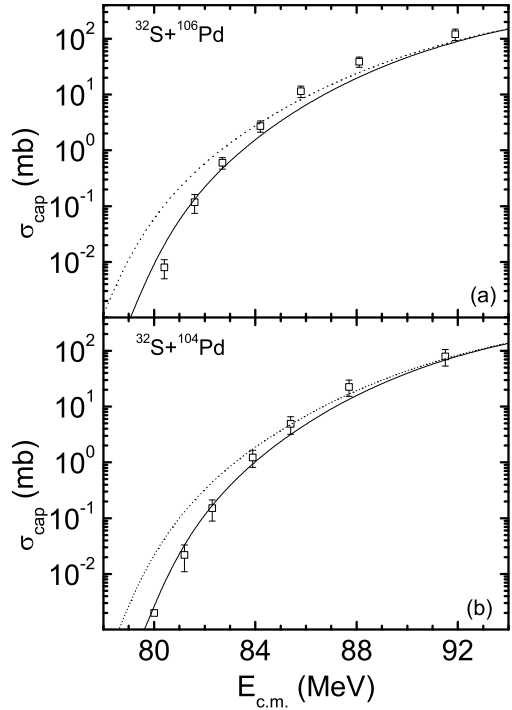}
\caption{The same as in Fig.~1, but for the  reactions $^{32}$S+$^{106}$Pd (a) and
$^{32}$S+$^{104}$Pd  (b). The experimental data are from Ref.~\protect\cite{Pengo}.
}
\label{3_fig}
\end{figure}
\begin{figure}[tbp]
\includegraphics[scale=1]{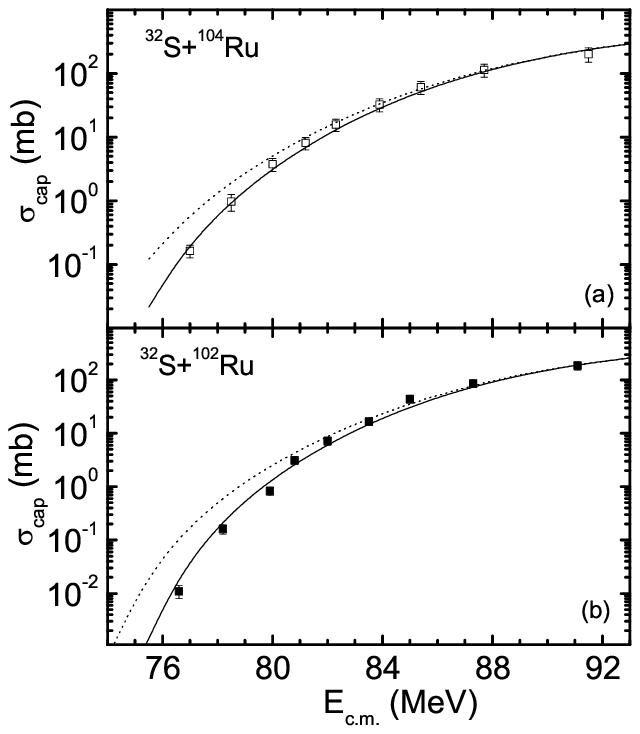}
\caption{The same as in Fig.~1, but   for the  reactions $^{32}$S+$^{104}$Ru (a) and
$^{32}$S+$^{102}$Ru  (b). The experimental data are from Ref.~\protect\cite{Pengo}.
}
\label{4_fig}
\end{figure}

Since $Q_{1n}<0$ in these reactions,
the enhancement arises not from the coherent
successive transfer of two single neutrons,
but from the direct transfer of one
spatially
correlated pair (the simultaneous transfer of two  neutrons).
%Neutron pair transfer indicates that stable nucleus  at low densities
%around nuclear surface may exhibit features of the
%strong coupling pairing, characterized by the small
%size  of the Cooper pair comparable with average internucleon distance.
%This may suggest a possible enhancement of the pair correlation
%in the surface of nuclei
%$^{48}$Ca, $^{64}$Ni, and $^{116,124,132}$Sn.
Our results show that the
capture (fusion) cross section of the reactions under consideration
can be described by assuming the preformed dineutron-like clusters in the
ground state of the nuclei $^{48}$Ca, $^{64}$Ni, and $^{116,124,132}$Sn.
Note that the strong spatial two-neutron correlation and the
strong surface enhancement of the neutron pairing  in the
cases of a slab, a semi-infinite nuclear matter, and the finite superfluid nuclei
are well known
and it is well established that nuclear superfluidity of the Cooper pairs
is mainly a
surface effect~\cite{past,Dean,Matsuo}.
\begin{figure}[tbp]
\includegraphics[scale=1]{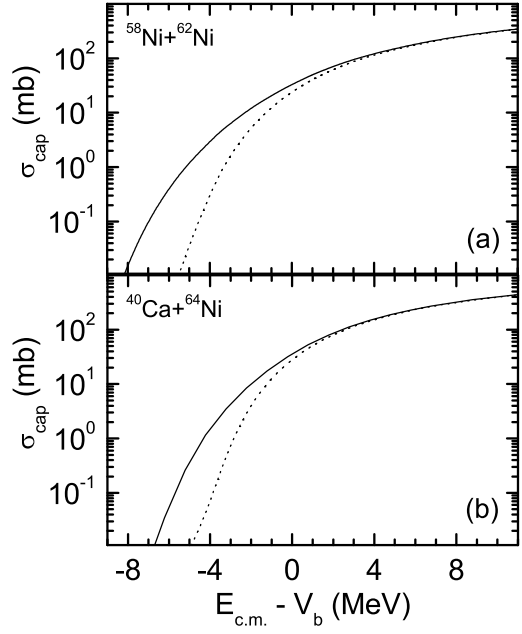}
\caption{The calculated capture cross section
vs $E_{\rm c.m.}-V_b$ for the reactions $^{58}$Ni+$^{62}$Ni (a)
$^{40}$Ca + $^{64}$Ni (b).
The results with and without taking into consideration
the neutron pair transfer  are shown by solid and dotted lines, respectively.
}
\label{5_fig}
\end{figure}
\begin{figure}[tbp]
\includegraphics[scale=1]{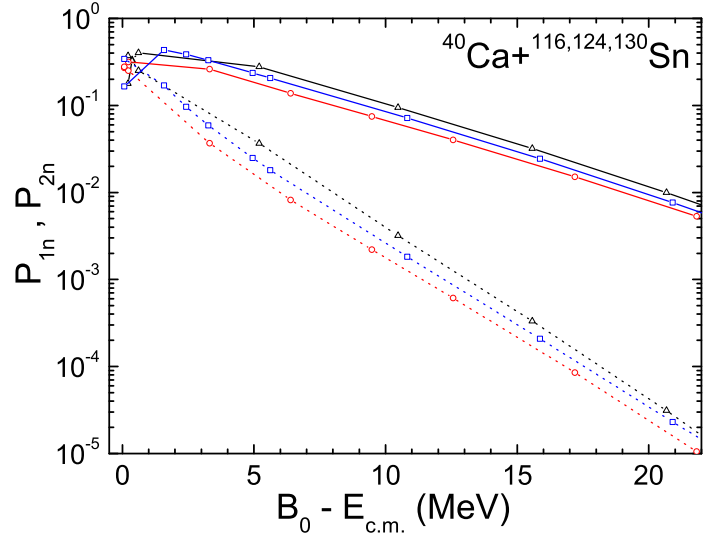}
\caption{(Color online) The calculated
one- (symbols connecting by  solid lines) and two-neutron (symbols connecting by dotted lines) transfer probabilities
vs $B_0-E_{\rm c.m.}$ for the reactions $^{40}$Ca+$^{116}$Sn (circles), $^{40}$Ca+$^{124}$Sn (triangles),
and $^{40}$Ca+$^{130}$Sn (squares).
}
\label{6_fig}
\end{figure}

%All systems in Figs.~2--6 show a steady decrease of the sub-barrier fusion cross sections
%with  pronounced change of slope.
%With decreasing $E_{\rm c.m.}$ up to about 3.5-5.0 MeV below
%the Coulomb barrier the regime of interaction is changed~\cite{EPJSub1,EPJSub2} because at external
%turning point the colliding nuclei do not reach the region of nuclear interaction
%where the friction plays a role.
%As a result, at smaller $E_{\rm c.m.}$ the cross sections fall with smaller rate.

Our calculations also show that the neutron pair transfer has to be taken into consideration in the
 description of the reactions $^{58}$Ni+$^{112,114,116,118,120}$Sn,
$^{32}$S+$^{94,96}$Mo,$^{100,102,104}$Ru,$^{104,106,108}$Pd, and
$^{18}$O+$^{112,118,124}$Sn (for example, see Figs.~3 and 4)~\cite{EPJSub2}.
In Figs. 3 and 4 one can see that
after  neutron pair transfer
in the reactions
$^{32}$S($\beta_2=0.312$)+$^{106}$Pd($\beta_2=0.229$)$\to ^{34}$S($\beta_2=0.252$)+$^{104}$Pd($\beta_2=0.209$),
$^{32}$S($\beta_2=0.312$)+$^{104}$Pd($\beta_2=0.209$)$\to ^{34}$S($\beta_2=0.252$)+$^{102}$Pd($\beta_2=0.196$)
or
$^{32}$S($\beta_2=0.312$)+$^{104}$Ru($\beta_2=0.271$)$\to ^{34}$S($\beta_2=0.252$)+$^{102}$Ru($\beta_2=0.24$),
$^{32}$S($\beta_2=0.312$)+$^{102}$Ru($\beta_2=0.24$)$\to ^{34}$S($\beta_2=0.252$)+$^{100}$Ru($\beta_2=0.215$)
the deformations of the nuclei  decrease  and
the values of the corresponding Coulomb barriers  increase and, respectively,
the capture cross sections decrease at the sub-barrier energies.
These results  indicate again the strong spatial two-neutron correlations
% and the
%strong surface enhancement of the neutron pairing
in the surface of the stable nuclei $^{18}$O,
$^{94,96}$Mo, $^{100,102,104}$Ru, $^{104,106,108}$Pd,
and
$^{112,114,116,118,120}$Sn. Since
the dominance of the dineutron-like clusters is found
in the surface of double magic, semimagic, and nonmagic nuclei,
one can conclude that this effect is  general for all stable and radioactive nuclei.
\begin{figure}[tbp]
\includegraphics[scale=0.5]{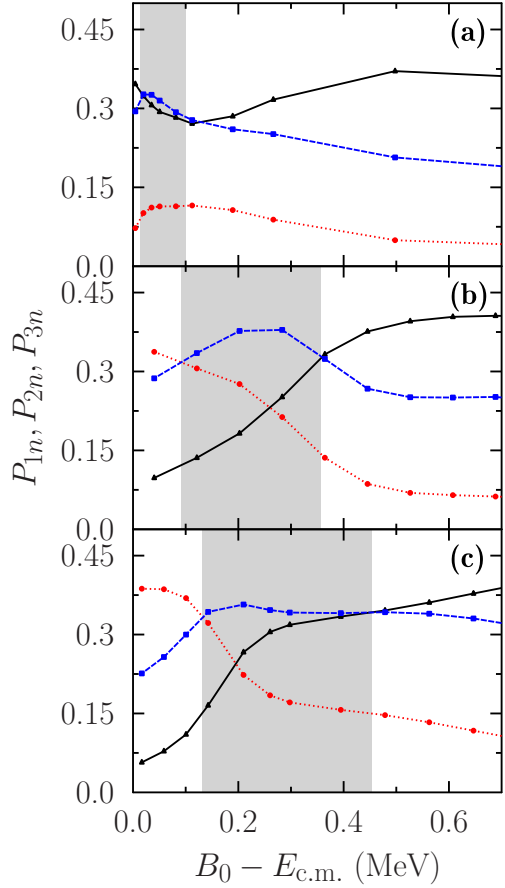}
\caption{(Color online) Focus on the calculated
one- (black filled triangles), two-(blue filled squares)  and three-(red filled circles) neutron transfer probabilities
as a function of $B_0-E_{\rm c.m.}$ for the reactions  $^{40}$Ca+$^{116}$Sn (a), $^{40}$Ca+$^{124}$Sn (b),
and $^{40}$Ca+$^{130}$Sn (c). In each case, the gray area indicates the energy region where the two-particle
channel dominates.
}
\label{7_fig}
\end{figure}
One can  make unambiguous statements
regarding the neutron pair transfer process in  the reactions
$^{40}$Ca + $^{62}$Ni ($Q_{1n}=-2.23$ MeV, $Q_{2n}=1.43$  MeV),
$^{40}$Ca + $^{64}$Ni ($Q_{1n}=-1.29$ MeV, $Q_{2n}=3.45$  MeV),
$^{40}$Ca + $^{114}$Sn ($Q_{1n}=-1.94$ MeV, $Q_{2n}=1.8$  MeV),
$^{40}$Ca + $^{118}$Sn ($Q_{1n}=-1.55$ MeV, $Q_{2n}=3.56$  MeV),
$^{40}$Ca + $^{120}$Sn ($Q_{1n}=-0.75$ MeV, $Q_{2n}=4.25$  MeV),
$^{40}$Ca + $^{122}$Sn ($Q_{1n}=-0.45$ MeV, $Q_{2n}=4.86$  MeV),
$^{58}$Ni + $^{62}$Ni ($Q_{1n}=-1.6$ MeV, $Q_{2n}=1.94$  MeV),
$^{60}$Ni + $^{64}$Ni ($Q_{1n}=-1.84$ MeV, $Q_{2n}=1.95$ MeV),
$^{64}$Ni + $^{128}$Sn ($Q_{1n}=-1.8$ MeV, $Q_{2n}=1.6$ MeV),
and
$^{64}$Ni + $^{130}$Sn ($Q_{1n}=-1.52$ MeV, $Q_{2n}=2.1$ MeV).
As  seen in Fig.~5, there is a considerable difference
between the sub-barrier capture cross sections with and without taking into consideration the
neutron pair transfer in these  reactions.
After  two-neutron transfer, the deformation of light nucleus strongly increases
and the capture cross section enhances.
The neutron pair  transfer induces the  effect of the quadrupole
deformation in the light nucleus.
The study of the capture reactions following the neutron transfer will provide
a good test for the effects of the neutron pair transfer.

\section{Neutron pair transfer phenomenon
in  heavy-ion sub-barrier reactions}
The Time-Dependent Hartree-Fock (TDHF) plus BCS approach \cite{Simenelnew,Scamps} has been recently used  \cite{Scamps} to extract the
one-, two-, three-neutrons transfer probabilities ($P_{1n}$, $P_{2n}$,  $P_{3n}$) in heavy-ion scattering reactions. It was shown that, when the energy
is well below the Coulomb barrier, the one-nucleon channel largely dominates. This is further illustrated here for the reactions
 $^{40}$Ca + $^{116,124,130}$Sn that have been discussed above and where the tin isotopes are superfluid.
In Fig. 6, the one- and two-neutron transfer probabilities are displayed as   functions of $B_0-E_{\rm c.m.}$ for the sub- and near-barrier binary collisions of $^{40}$Ca and tin isotopes. The Coulomb barrier (capture threshold energy) $B_0$ is deduced from the mean-field transport theory. This barrier are equal to $116.41\pm 0.07$ ($^{116}$Sn),  $114.69\pm0.04$ ($^{124}$Sn) and $113.92\pm0.02$
($^{130}$Sn) MeV. It was found that the calculated $B_0$ are insensitive to the introduction of pairing and in a good agreement with the barriers extracted from the experimental data \cite{Scamps}. Note that the presented calculation are shown for the mixed pairing interaction only. The use of other interaction (surface or volume) leads to similar conclusions.
%The pairing interaction strength is chosen as follows:
%\begin{eqnarray}
%V_\tau=V_0^{\tau \tau}(1-\eta\rho([{\bf r}+{\bf r'}]/2)/\rho_0)\delta_{\bf rr'}[1-P_{\sigma\sigma'}],
%\end{eqnarray}
%where $P_{\sigma\sigma'}$ is the spin exchange operator,
%$V_0^{nn}=...$ MeV$\times$fm$^3$ for neutrons, $V_0^{pp}=...$ MeV$\times$fm$^3$ for protons, $\rho_0=0.16$ fm$^{-3}$,  and
%$\tau=n,p$.
%The results of calculations show that the extracted transfer probabilities are insensitive to the type of effective pairing interaction
%(volume [$\eta=0$], surface [$\eta=1$], mixed [$\eta=0.5$]) when the interaction is adjusted to reproduce the experimental pairing
%gap \cite{Scamps}.
%Since the all types of pairing interaction lead to the same transfer probabilities,  only results with the mixed interaction are
%presented below.
Figure 6 gives an interesting insight in the one- and two-neutron transfers. As seen, a strong enhancement of $P_{1n}$ and $P_{2n}$ occurs with
increasing bombarding energy. Since the enhancement of $P_{2n}$ is stronger than that of $P_{1n}$, these probabilities become close to each other with
decreasing $B_0-E_{\rm c.m.}$.
This is indeed observed experimentally in Refs. \cite{Corradi,Corradi2,Simenel}  where it was found that $P_{2n}$ grows faster than  $P_{1n}$ with decreasing $B_0-E_{\rm c.m.}$ at energy relatively far below the Coulomb barrier.

In Fig. \ref{7_fig}, a closer look is made on the one-, two- and three-neutrons transfer
channels at the vicinity of the Coulomb barrier for the different tin isotopes. In all cases, as the energy approaches the capture barrier
energy, there exist an energy range where $P_{2n} > P_{1n}$ dominates (shaded area). We also note that the energy windows where
the two-nucleon channel becomes dominant increases as the neutron nucleus become more exotic.

This   evidently supports  our assumption about important role of the two-neutron transfer (compared to the one-neutron transfer)
in the capture process, because in
the TDHF calculation the scattering trajectory of two heavy ions at energy near the Coulomb barrier is close to the capture trajectory. Note that in the capture process
the system trajectory crosses the  barrier position $R=R_b$ at any energies.
The results of our calculations
predict that there is the crossing point of $P_{2n}$ and $P_{1n}$  at energy very close to the
Coulomb barrier.
Just before reaching $R_b$ the neutron-pair transfer becomes the dominant channel. Thus,
our assumption about two-neutron transfer before the capture is correct. The transfer more than two neutrons mainly
occurs at $R<R_b$, i.e., just after the capture.

\section{Summary}
Within the quantum diffusion approach
it turns out that the sub-barrier capture (fusion) reactions with $Q_{1n}<0$ and $Q_{2n}>0$ may help  us understanding
of the neutron pair transfer  and of the pair correlation phenomenon on the surface of a nucleus.
In these reactions the main contribution to transfer
is due to the dineutron-like cluster component.
In the capture process, the transfer of  neutron pair before the crossing of the
Coulomb barrier is a clear signature
of the strong correlations between the transferred nucleons
%affecting the dynamics
and the surface  character of  pairing interaction.
Our results  indicate the dominance of the dineutron structure
(of the preformed dineutron-like  clusters)
in the surface of the stable and unstable nuclei $^{18}$O, $^{48}$Ca, $^{64}$Ni,
$^{94,96}$Mo, $^{100,102,104}$Ru, $^{104,106,108}$Pd,
and
$^{112,114,116,118,120,124,132}$Sn.
Measurements of sub-barrier capture cross sections in various reactions can be
utilized to study the role of pairing correlations between the transferred nucleons.
The information obtained from the sub-barrier
capture (fusion) reactions is complementary to
that obtained from the two-neutron transfer reactions
such as ($p$,$t$) or ($t$,$p$) and the multinucleon transfer reactions.
%We have shown that the information
%obtained from the sub-barrier capture (fusion) reactions is complementary to
%that obtained from the two-neutron transfer reactions such as ($p$,$t$) or ($t$,$p$).
%
%
%In the area of nuclear structure, a neutron pair transfer can be used for the study of neutron-neutron correlation
% in nuclei and nuclear Josephson effect [].

%The pairing interaction induced by the coupling to the phonons is expected to increase the surface enhancement.

%The matrix element of the pair transfer has direct relation to the degree of the pair correlation.

%In the extreme case of infinitely strong correlation the one-step mechanism is expected to prevail,
%the two-step sequential process being instead dominant in the case of pure uncorrelated nucleons.

%The interplay of these two processes is one of the key point in the understanding of ...

%The enhancement is supposed to further increase if both nuclei are in the superfluid case (Josephson effect),
%as can be the case of transfer reactions induced be heavy ions.
%The tin isotopes with closed proton shell ($Z=50$) and open neutron shells are considered to represent the
%ideal case of a neutron pairing superfluid.

Employing the Time-Dependent Hartree-Fock plus BCS approach \cite{Scamps}, we demonstrated
the important role of two-neutron transfer channel in the heavy-ion scattering
 at sub-barrier energies close to the Coulomb barrier. We suggest
the experiments $^{40}$Ca + $^{116,124}$Sn and $^{40}$Ca + $^{48}$Ca to check our predictions.

We thank R.V.~Jolos  and H.Q.~Zhang for fruitful discussions and  suggestions.
%We are grateful to J.F.~Liang for providing us his experimental data.
This work was supported by DFG  and RFBR (grants 12-02-31355, 13-02-12168, 13-02-000080, 12-02-91159).
The IN2P3(France)-JINR(Dubna)
%, MTA(Hungary)-JINR(Dubna)
and Polish - JINR(Dubna)
Cooperation Programmes are gratefully acknowledged.\\

%\newpage


\begin{thebibliography}{99}
\bibitem{BohrNathan} O.~Nathan and A.~Bohr, in {\it Int. Symp. on Nuclear Structure} (Dubna, 1968).
\bibitem{vonOertzen} W.~von~Oertzen and A.~Vitturi, Rep. Prog. Phys. {\bf 64}, 1247 (2001).
\bibitem{Broglia} D.R.~Bes and R.A.~Broglia, Nucl. Phys. {\bf 80}, 289 (1966);
 R.A.~Broglia, O.~Hansen, and C.~Riedel, Adv. Nucl. Phys. {\bf 6}, 287 (1973).
\bibitem{Kart} R.V.~Jolos, V.G.~Kartavenko, F.~D\"onau, and D.~Janssen, Theor. Math. Fys.
{\bf 14}, 70 (1973); R.V.~Jolos and V.G.~Kartavenko, Yadernaya Fysika {\bf 19}, 964 (1974);
R.V.~Jolos, V.G.~Kartavenko, and V.~Rybarska, Theor. Math. Phys.
 {\bf 20}, 353 (1974); R.V.~Jolos, V.G.~Kartavenko, and S.M.~Semenov,
Yadernaya Fysika  {\bf 22}, 1121 (1975).
\bibitem{past} R.H.~Ibarra, N.~Austern, M.~Vallieres, and D.H.~Feng, Nucl. Phys. {\bf A288}, 397 (1977);
F.~Catara, A.~Insolia, E.~Maglione, and A.~Vitturi, Phys. Rev. C {\bf 29}, 1091 (1984);
L.~Ferreira, R.~Liotta, E.H.~Dasso, R.A.~Broglia, and A.~Winther, Nucl. Phys. {\bf A426}, 276 (1984);
M.A.~Tischler, A.~Tonina,  and G.G.~Dussel, Phys. Rev. C {\bf 58}, 2591 (1998).
\bibitem{Corradi} L.~Corradi, G.~Pollarolo, and S.~Szilner,
J. Phys. G  {\bf 36}, 113101  (2009).
\bibitem{Corradi2} L.~Corradi {\it et al.},
Phys. Rev. C {\bf 84}, 034603 (2011).
\bibitem{Simenel} M.~Evers, M.~Dasgupta, D.J.~Hinde, D.H.~Luong, R.~Rafiei, R.~du~Rietz, and C.~Simenel,
Phys. Rev. C {\bf 84}, 054614 (2011).

\bibitem{Marques} F. M. Marqu\'es, {\it  et  al.},
Phys. Rev. C {\bf 64}, 061301(R) (2001).
%; F. M. Marqu\'es, {\it  et  al.},
%Phys. Rev. C {\bf 65}, 044006 (2002).
\bibitem{Kolata} J.J.~Kolata {\it  et  al.},
Phys. Rev. C {\bf 75}, 031302 (2007).
\bibitem{Ershov} S.N.~Ershov and B.V.~Danilin,  Particles and Nuclei  {\bf 39}, 1623 (2008).
%\bibitem{Simpson} E.C.~Simpson, J.A.~Tostevin, D.~Bazin, B.A.~Brown, and A.~Gade,
% Phys. Rev. Lett. {\bf 102}, 132502 (2009).
\bibitem{Lacroix0} M.~Assi\'e and D.~Lacroix,  Phys. Rev. Lett. {\bf 102}, 202501 (2009);
J.A.~Scarpaci {\it  et  al.}, Phys. Rev. C {\bf 82}, 031301(R) (2010).
\bibitem{Grigor} L.V.~Grigorenko, I.G.~Mukha, C.~Scheidenberger, and M.V.~Zhukov, Phys. Rev. C {\bf 84}, 021303(R) (2011).
\bibitem{Spyrou} A.~Spyrou {\it  et  al.},  Phys. Rev. Lett. {\bf 108}, 102501 (2012).
%
\bibitem{Fortunato} L.~Fortunato, W.~von~Oertzen, H.M.~Sofia, and A.~Vitturi, Eur. Phys. J. A {\bf 14}, 37 (2002).
\bibitem{Volya} A.~Volya, V.~Zelevinsky, and B.A.~Brown,
Phys. Rev. C {\bf 65}, 054312 (2002).
\bibitem{Dean} D.J.~Dean and M.~Hjorth-Jensen, Rev. Mod. Phys. {\bf 75}, 607 (2003).
\bibitem{Khan} E.~Khan, N.~Sandulescu, N.V.~Giai, and M.~Grasso, Phys. Rev. C {\bf 69}, 014314 (2004).
\bibitem{Saper1} M.~Baldo, U.~Lombardo, E.E.~Saperstein, and M.V.~Zverev, Phys. Rep. {\bf 391}, 261 (2004).
%\bibitem{Hag05} K. K. Hagino and H. Sagawa, Phys. Rev. C 72, 044321(2005).
%\bibitem{Avez} B.~Avez, S.~Cimenel, and Ph.~Chomas, Phys. Rev. C {\bf 78}, 044318 (2008).
\bibitem{Matsuo}  M.~Matsuo, K.Mizuyama, and Y.~Serizawa, Phys. Rev. C {\bf 71}, 064326 (2005);
M.~Matsuo and Y.~Serizawa, Phys. Rev. C {\bf 82}, 024318 (2010);
H.~Shimoyama and M.~Matsuo, Phys. Rev. C {\bf 84}, 044317 (2011).
%; M.~Matsuo, in {\it Fifty years of nuclear BCS,
%World Scientific (2012)}, eds. R.A.~Broglia and V.~Zelevinsky.
%\bibitem{Lacroix2} G.~Hupin and D.~Lacroix, , Phys. Rev. C {\bf 83}, 024317 (2011).
\bibitem{Grasso} E.~Pllumbi, M.~Grasso, D.~Beaumel, E.~Khan, J.~Margueron, and J.~van~de~Wiele,
Phys. Rev. C {\bf 83}, 034613 (2011).
\bibitem{Broglia2} G.~Potel, F.~Barranco, F.~Marini, A.~Idini,
E.~Vigezzi, and R.A.~Broglia
Phys. Rev. Lett. {\bf 107}, 092501 (2011).
\bibitem{Lacroix} M.~Grasso, D.~Lacroix, and A.~Vitturi, Phys. Rev. C {\bf  85}, 034317 (2012);
\bibitem{Lacroix2}  G.~Scamps, D.~Lacroix, G.F. Bertsch, and K. Washiyama,
 Phys. Rev. C {\bf  85}, 034328 (2012).
\bibitem{Lacroix3} D. Gambacurta and D. Lacroix, Phys. Rev. C {\bf  86}, 064320 (2012).
\bibitem{Sambataro} M.~Sambataro, Phys. Rev. C {\bf  85}, 064326 (2012).
%
\bibitem{Zhukov} M.V.~Zhukov, B.V.~Danilin, D.V.~Fedorov, J.M.~Bang, I.J.~Thompson, and J.S.~Vaagen,
Phys. Rep.  {\bf  231}, 151 (1993).
\bibitem{Mar} F. M. Marqu\'es, {\it  et  al.},
Phys. Lett. B {\bf 476}, 219 (2000).
\bibitem{Barranco} F.~Barranco,
P.F. Bortignon,
R.A. Broglia,
G. Col\'o, and
E. Vigezzi , Eur. Phys. J. A {\bf 11}, 385 (2001);
E.~Vigezzi, F. Barranco, R.A. Broglia, G. Col\'o, G. Gori, and F. Ramponi,
Nucl. Phys.  {\bf A752}, 600 (2005).
%\bibitem{Au} T.~Aumann, Eur. Phys. J. A {\bf 26}, 441 (2005).
\bibitem{Mat} M.~Matsuo, Phys. Rev. C {\bf 73}, 044309 (2006).


%
\bibitem{Ter-Akopian} G.M.~Ter-Akopian {\it  et  al.}, Phys. Lett. B {\bf 426}, 251 (1998).
\bibitem{DeYoung} P.A.~DeYoung {\it  et  al.}, Phys. Rev. C {\bf 71}, 051601 (2005).
\bibitem{Nakamura} T.~Nakamura {\it  et  al.}, Phys. Rev. Lett. {\bf 96}, 252502 (2006).
\bibitem{Moeller} P.~Mueller {\it  et  al.}, Phys. Rev. Lett. {\bf 99}, 252501 (2007).
\bibitem{Chatterjee} A.~Chatterjee {\it  et  al.}, Phys. Rev. Lett. {\bf 101}, 032701 (2008).%
\bibitem{Saper2} E.E.~Saperstein and M.V.~Troitsky, Yad. Fiz. {\bf 1}, 10 (1965).
\bibitem{Hag05} K. Hagino and H. Sagawa, Phys. Rev. C 72, 044321(2005).
%
\bibitem{Dobaczewski} J.~Dobaczewski, W.~Nazarewicz, T.R.~Werner, J.F.~Berger, C.R.~Chinn, and J.~Decharg\'e ,
Phys. Rev. C {\bf 53}, 2809 (1996).
\bibitem{Gra13} M. Grasso, Phys. Rev. C 87, 064308 (2013)
\bibitem{Schuck} N.~Pillet, N.~Sandulescu,  and P.~Schuck, Phys. Rev. C {\bf 76}, 024310 (2007);
N.~Pillet, N.~Sandulescu, P.~Schuck, and J.-F.Berger, Phys. Rev. C {\bf 81}, 034307 (2010);
X.~Vi\~nas, P.~Schuck, and N.~Pillet, Phys. Rev. C {\bf 82}, 034314 (2010).
\bibitem{EPJSub}        V.V.~Sargsyan, G.G.~Adamian, N.V.~Antonenko, and W.~Scheid, Eur. Phys. J. A {\bf 45}, 125 (2010).
\bibitem{EPJSub1}        V.V.~Sargsyan, G.G.~Adamian, N.V.~Antonenko,  W.~Scheid, and H.Q.~Zhang,
Eur. Phys. J. A {\bf 47}, 38 (2011);
%\bibitem{Conf}      V.V.~Sargsyan, G.G.~Adamian, N.V.~Antonenko,  W.~Scheid, and H.Q.~Zhang,
J. of Phys.: Conf. Ser. {\bf 282}, 012001 (2011);  EPJ Web Conf. {\bf 17}, 04003 (2011).
\bibitem{EPJSub2}        V.V.~Sargsyan, G.G.~Adamian, N.V.~Antonenko,  W.~Scheid, and H.Q.~Zhang,
Phys. Phys. C {\bf 84}, 064614 (2011);  Phys. Rev. C  {\bf 85}, 024616 (2012); Phys. Rev. C {\bf 85}, 069903(E) (2012).
\bibitem{EPJSub3}        V.V.~Sargsyan, G.G.~Adamian, N.V.~Antonenko,  W.~Scheid, C.J.~Lin, and H.Q.~Zhang,
Phys. Phys. C {\bf 85},  017603 (2012);
Phys. Phys. C {\bf 85},  037602 (2012); R.A.~Kuzyakin,
V.V.~Sargsyan, G.G.~Adamian, N.V.~Antonenko,  E.E.~Saperstein, and S.V.~Tolokonnikov,
Phys. Rev. C {\bf 85},  034612 (2012).

\bibitem{Ayik333}       S.~Ayik, B.~Yilmaz, and D.~Lacroix,  Phys. Rev. C {\bf 81}, 034605 (2010).
\bibitem{our}           V.V.~Sargsyan, Z.~Kanokov, G.G.~Adamian, N.V.~Antonenko, and W.~Scheid,
                        Phys. Rev. C {\bf 80}, 034606 (2009); Phys. Rev. C {\bf 80}, 047603 (2009);
                        V.V.~Sargsyan, Z.~Kanokov, G.G.~Adamian, and N.V.~Antonenko, Part. Nucl. {\bf 41}, 175 (2010).
%
%\bibitem{VAZ}           G.G.~Adamian, N.V.~Antonenko, Z.~Kanokov, and V.V.~Sargsyan,
%                        Teor. Mat. Fiz. {\bf 145}, 87 (2005)
%                        [Theor. Math. Phys. {\bf 145}, 1443 (2006)];
%                        Z.~Kanokov, Yu.V.~Palchikov,  G.G.~Adamian, N.V.~Antonenko, and W.~Scheid, Phys. Rev. E {\bf %71}, 016121 (2005).
%                        Yu.V.~Palchikov, Z.~Kanokov,  G.G.~Adamian, N.V.~Antonenko, and W.~Scheid, Phys. Rev. E {\bf %71}, 016122 (2005).
\bibitem{SSzilner}   S.~Szilner  {\it et al.}, Phys. Rev. C 84, 014325 (2011).
\bibitem{Dasso} R.A.~Broglia, C.H.~Dasso,
S.~Landowne, and A.~Winther, Phys. Rev. C {\bf 27}, 2433 (1983);
R.A.~Broglia, C.H.~Dasso, S.~Landowne, and  G.~Pollarolo, Phys. Lett. B  {\bf 133}, 34 (1983);
C.H.~Dasso, S.~Landowne, and A.~Winther, Nucl. Phys.  {\bf A405}, 381 (1983).
\bibitem{Hofman}        H.~Hofmann, Phys. Rep.  {\bf 284}, 137 (1997).
\bibitem{Ayik}  L.~F.~Canto, Nucl. Phys.  {\bf A491}, 337 (1989);
N.~Takigawa, S.~Ayik, K.~Washiyama, and S.~Kimura, Phys. Rev. C {\bf 69}, 054605 (2004);
S.~Ayik, B.~Yilmaz, A.~Gokalp, O.~Yilmaz, and N.~Takigawa, Phys. Rev. C {\bf 71}, 054611 (2005).
\bibitem{Hupin} G.~Hupin and D.~Lacroix,  Phys. Rev. C {\bf 81}, 014609 (2010).
\bibitem{AAShobzor} G.G.~Adamian, N.V.~Antonenko, and W.~Scheid, {\it Clusters in Nuclei}, Vol.2, C.
Beck (ed.),    Lecture Notes in Physics {\bf 848}, p. 165, Springer-Verlag, Berlin,
Heidelberg (2012).
\bibitem{Ram}           S.~Raman, C.W.~Nestor,~Jr, and P.~Tikkanen,
                        At. Data Nucl. Data Tables {\bf 78}, 1 (2001).
%
\bibitem{trotta40ca48ca}    H.A.~Aljuwair, R.J. Ledoux, M. Beckerman, S.B. Gazes,
J. Wiggins, E. R. Cosman, R.R. Betts, S. Saini, and  Ole Hansen,    Phys. Rev. C {\bf 30}, 1223 (1984);
C.L.~Jiang~{\it et al.}, Phys. Rev. C {\bf 82},  041601(R) (2010).
%G.~Montagnoli and A.M.~Stefanini, EPJ Web Conf. {\bf 17}, 05001 (2011).
%M.~Trotta~{\it et al.}, Phys. Rev. C {\bf 65}, 011601R (2001).
\bibitem{Stefanini40ca116124sn}
%A.M.~Stefanini~{\it et al.},   Eur. Phys. J. A {\bf 23}, 1401 (1997);
                                F.~Scarlassara, , S. Beghini, G. Montagnoli, G.F. Segato, D. Ackermann, L. Corradi, C.J. Lin, A.M. Stefanini, and L.F. Zheng, Nucl. Phys.  {\bf A672}, 99 (2000).
\bibitem{LiangNi64Sn132}        J.F.~Liang, D. Shapira, C.J. Gross, R.L. Varner, J.R. Beene, P.E. Mueller,
and D.W. Stracener , Phys. Rev. C {\bf 78}, 047601 (2008).
\bibitem{Beckerman58Ni5864Ni74Ge}     M.~Beckerman, M.~Salomaa, A.~Sperduto,
J.D.~Molitoris, and A.~DiRienzo, Phys. Rev. C {\bf 25},  837 (1982).
\bibitem{Pengo} R.~Pengo, D. Evers, K.E.G. L\''obner, U. Quade, K. Rudolph, S.J. Skorka, and I. Weidl, Nucl. Phys. {\bf A411}, 255 (1983).
\bibitem{Stef} A.M.~Stefanini~{\it et al.},   Nucl. Phys.  {\bf A456}, 509 (1986);
A.M.~Borges, C.P.~da~Silva, D.~Pereira, L.C.~Chamon, E.S.~Rossi, Jr., and C.E.~Aguiar,
Phys. Rev. C {\bf 46}, 2360 (1992); S.~Kalkal {\it et al.}, Phys. Rev. C {\bf 83}, 054607 (2011).
%\bibitem{newpaper} V.V.~Sargsyan, G.G.~Adamian, N.V.~Antonenko,  W.~Scheid, and H.Q.~Zhang,
%Phys. Rev. C {\bf 85} (2012) in print.
%%%

\bibitem{Simenelnew} C.~Simenel   and B.~Avez, Int. J. Mod. Phys. E {\bf  17}, 31 (2008);
C.~Simenel, {\it Clusters in Nuclei}, Vol.3, C.
Beck (ed.), Lecture Notes in Physics {\bf 875}, p. 95, Springer-Verlag, Berlin,
Heidelberg   (2013).

\bibitem{Scamps} G.~Scamps and D.~Lacroix, Phys. Rev. C {\bf  87}, 014605 (2013).

\end{thebibliography}
\end{document}